\documentclass[pre,twocolumn,floatfix,amsmath,amsfonts,longbibliography]{revtex4}
\usepackage{graphicx,epstopdf,url}
\usepackage{color}
\usepackage[colorlinks=true,urlcolor=blue,citecolor=blue,linkcolor=blue,urlcolor=blue]{hyperref}
\usepackage[active]{srcltx}

\begin{document}


\title{Structural and helix reversal defects of carbon nanosprings}

\author{Alexander V. Savin}
\email[]{asavin@chph.ras.ru}
\affiliation{Semenov Institute of Chemical Physics, Russian Academy of Sciences, Moscow 119991, Russia}
\affiliation{Plekhanov Russian University of Economics, Moscow 117997, Russia}

\author{Elena A. Korznikova}
\email[]{elena.a.korznikova@gmail.com}
\affiliation{Ufa University of Science and Technology, Zaki Validi St. 32, 450076 Ufa, Russia}
\affiliation{Polytechnic Institute (Branch) in Mirny, North-Eastern Federal University, Tikhonova St. 5/1, 678170 Mirny, Sakha Republic (Yakutia), Russia}

\author{Sergey~V.~Dmitriev}
\email[]{dmitriev.sergey.v@gmail.com}
\affiliation{Ufa State Petroleum Technological University, Kosmonavtov st. 1, Ufa 450064, Russia}

\begin{abstract}
Due to their chiral structure, carbon nanosprings possess unique properties that are promising for nanotechnology applications.
The structural transformations of carbon nanosprings in the form of spiral macromolecules derived from planar coronene and kekulene molecules (graphene helicoids and spiral nanoribbons) are analyzed using molecular dynamics simulations.
While the tension/compression of such nanosprings has been analyzed in the literature, this study investigates other modes of deformation, including bending and twisting.
Depending on the geometric characteristics of the carbon nanosprings, the formation of structural and helix reversal defects is described. It is found that nanosprings demonstrate a significantly higher coefficient of axial thermal expansion than many metals and alloys.
These results are useful for designing nanosensors that operate over a wide temperature range.

\end{abstract}

\maketitle

\section{Introduction \label{s1}}

Carbon, an element in the fourth group of the periodic table, can form a wide variety of $sp^2$ structures, including chiral structures that cannot be superimposed on their mirror images.
Among them are the carbon microcoils/nanocoils~\cite{Yang2004,Qian2019},
coiled carbon nanotubes~\cite{Ghaderi2012,Poggi2004,Liu2013}, carbon nanocones~\cite{Tang2012}, as well as the macromolecules composed of helicene and kekulene molecules~\cite{Nakakuki2018,Nakakuki2018a,Savin2024}.
The helical two-dimensional materials can be grown on nonplanar substrates~\cite{Zhao2020}.
The geometry of the coiled carbon nanotubes is controlled by the distribution of Stone-Wales and vacancy defects~\cite{Bie2021,Bie2023,Sharifian2025}.
The helical graphene is actually a helical polymer~\cite{Nakano2001,Yashima2009,Ikai2023,Qiu2025}  based on helicene and kekulene molecules.

The helical graphene displays fascinating electronic properties~\cite{Avdoshenko2013,Tan2018,Zhang2014,Atanasov2015}, which are primarily determined by the interactions between layers~\cite{Xu2017,Korhonen2014}.
The twisted nanoribbons can be used as nanocoils of inductance~\cite{Xu2016,Porsev2023}.
Coiled carbon nanotubes have potential in applications as nanoelectromechanical devices~\cite{Lin2023}.
A change in the pitch of the helical graphene results in the metal-semiconductor transition~\cite{Liu2019a}.
The double-layer spiral graphene is metallic in an equilibrium state~\cite{Thakur2021}.
It has been determined that the electronic properties of helical graphene with armchair and zigzag edges differ significantly~\cite{Zhou2022}.

Helical carbon nanosprings are a unique material that combines the properties of single-layer nanoribbons and multilayer graphene.
Due to their distinctive properties, they are of significant interest for nanotechnology.
Not only electronic but also magnetic properties of helical graphene depend on the structure and elastic stretching~\cite{Porsev2019}.
Compression and stretching of helical graphene nanoribbons have been shown to cause a significant change in their thermal conductivity~\cite{Zhan2018,Norouzi2020,Sharifian2022,Li2023}.

The mechanical properties of helical graphene nanoribbons have been analyzed in numerous studies.
The nanosprings exhibit exceptional elasticity, with a maximum reversible tensile strain of hundreds percent~\cite{Sestak2015,Zhan2017}.
The nanosprings, under tension, demonstrate a deviation from Hooke's law due to the breaking of the van der Waals bonds between coils, which leads to non-homogeneous stretching~\cite{Zhan2018a}.
Graphene nanoribbon nanosprings exhibit a distinctive force-strain relationship under tension, characterized by a constant force plateau across a broad range of tensile strain~\cite{Savin2024,Norouzi2019,Zhu2021,Sharifian2019}.
This phenomenon was explained in~\cite{Savin2024} by non-convex dependence of the potential energy of a structural unit of the nanospring as the function of tensile strain.
A similar phenomenon has been observed for DNA~\cite{Smith1996,Savin2013,Afanasyev2022} and for intermetallic NiAl and FeAl nanofilms~\cite{Babicheva2013,Babicheva2013a,Bukreeva2013}.

A comprehensive analysis of the mechanical and thermal properties of carbon nanosprings has been conducted, with a focus on tension and compression deformation~\cite{Zhan2018a,Dmitriev2012,Liu2019,Savin2024,Li2023,Mokhalingam2022}.
Other modes of loading have not been addressed by the researchers, and here axial compression, bending, and twisting of the carbon nanosprings is analyzed using molecular dynamics simulations.
The structures under consideration in this work can be divided into two categories: helical graphene nanoribbons (see Fig.~\ref{fig01}(a,c)) and helicoids (see Fig.~\ref{fig01}(b,d)).
The chiral structures of the first type possess an inner channel, while the structures of the second type are devoid of such a feature.

In this work, we show that helical carbon nanosprings, in addition to their high tensile ability, have other unique mechanical properties.
Thus, their compression can lead to the formation of stable folded structures with fractures, and their twisting can lead to the formation of structures with localized helix reversal defects separating parts of the nanospring with opposite chirality.
Carbon nanosprings exhibit remarkable mechanical and electronic properties. 
However, their performance is highly sensitive to structural and helix reversal defects, which can alter elasticity, conductivity, and stability. 
The work will provide a detailed analysis of these defects.

\section{Model  \label{s2}}
\label{Model}

\begin{figure}[tb]
\begin{center}
\includegraphics[angle=0, width=1.0\linewidth]{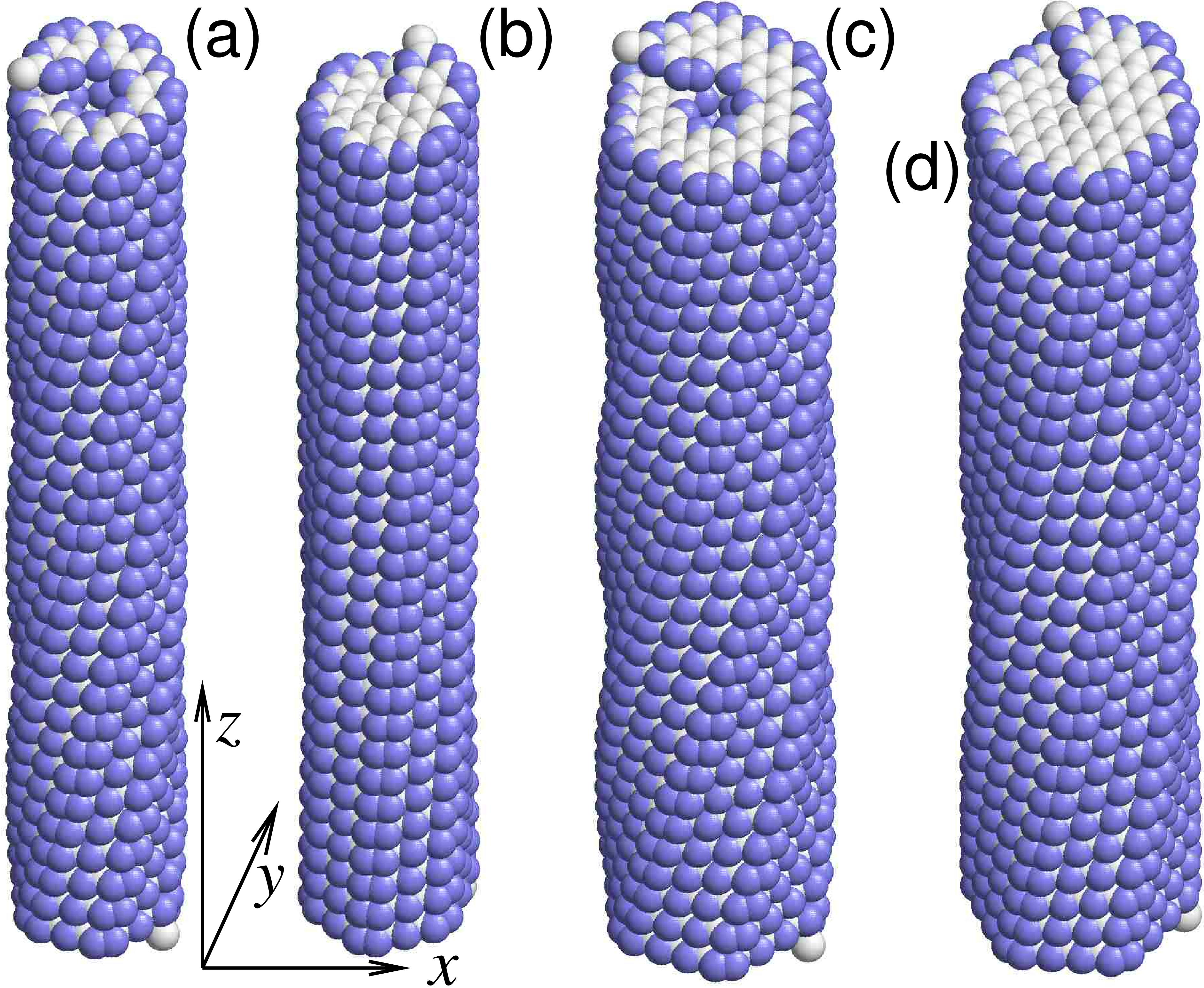}
\end{center}
\caption{\label{fig01}\protect
Spiral $l$-kekulene  graphene nanoribbon (C$_{l^2-1}$H$_{l+1}$)$_{180}$ constructed from the $l$-kekulene molecule C$_{6(l^2-1)}$H$_{6(l+1)}$: (a) $l=3$ (helix kekulene); (c) $l=4$ (helix circumkekulene).
Graphene helicoid (helical $l$-helicene) (C$_{l^2}$H$_l$)$_{180}$ built from $l$-coronene molecule C$_{6l^2}$H$_{6l}$: (b) $l=3$ (helix circumhelicene); (d) $l=4$ (helix dicircumhelicene).
The united CH atoms are shown in blue and the inner carbon atoms are shown in light gray. The nanosprings in (a) and (c) have an inner channel. Those in (b) and (d), however, do not.
}
\end{figure}

Consider helical molecular structures derived from planar molecules $l$-kekulene C$_{6(l^2-1)}$H$_{l+1}$ ($l\ge 3$) and $l$-coronene C$_{6l^2}$H$_{6l}$ ($l\ge 2$) lying in the $xy$ plane, by cutting them along the radius and further spiral extension along the $z$ axis -- see Fig.~\ref{fig01}(a,c) and (b,d).
The ground homogeneous state of such helical structures can be represented as successive shifts by $\Delta z$ and rotations by angle $\Delta\phi\approx \pi/3$ around the $z$-axis of a monomer of $N_{\rm C}$ and $N_{\rm H}$ carbon and hydrogen atoms (for $l$-kekulene $N_{\rm C}=l^2-1$, $N_{\rm H}=l+1$, for $l$-coronene $N_{\rm C}=l^2$, $N_{\rm H}=l$).
When viewed from above, such helical structures will look like flat $l$-kekulene and $l$-coronene molecules.

To simplify the modeling, valence-bonded CH groups of atoms at the edges of spiral structures are considered as a single carbon atom of mass $M_1=13m_p$, while all other inner carbon atoms have the mass $M_0=12m_p$, where $m_p=1.6601\times 10^{-27}$~kg is the proton mass.
In this approach, each cell of the helix will consist of only $N_{\rm C}-N_{\rm H}$ carbon atoms of mass $M_0$ and $N_{\rm H}$ united atoms of mass $M_1$.
In Fig. \ref{fig01} the united atoms are shown in blue and the inner atoms are shown in light gray.

The coordinates of the carbon atoms of the $n$-th cell of the helix are completely determined by the by the coordinates of the atoms of the previous $n-1$ cell:
\begin{eqnarray}
x_{n,j,1}&=&x_{n-1,j,1}\cos(\Delta\phi)-x_{n-1,j,2}\sin(\Delta\phi), \nonumber\\
x_{n,j,2}&=&x_{n-1,j,1}\sin(\Delta\phi)+x_{n-1,j,2}\cos(\Delta\phi), \label{f1}\\
x_{n,j,3}&=&x_{n-1,j,3}+\Delta z,~j=1,...,N_{\rm C}, \nonumber
\end{eqnarray}
where the vector ${\bf x}_{n,j}=(x_{n,j,1},x_{n,j,2},x_{n,j,3})$ defines the coordinates of the $j$th atom of the $n$th unit cell, $N$ is the number of unit cells (the length of the spiral $L=(N-1)\Delta z$).
The longitudinal pitch is $\Delta z \approx 0.58$~AA, the angular pitch of the spiral is $\Delta\phi\approx 61^\circ$ \cite{Savin2024}.

A helical nanospring built from $l$-coronene  molecules (helical $l$-helicene) has the chemical formula (C$_{l^2}$H$_l$)$_N$, where $N$ is the number of structural units, $l\ge 2$.
Such a $l$-coronene nanospring has the shape of a graphene helicoid, see Fig.~\ref{fig01} (b) and (d).
A spiral nanospring built from $l$-kekulene molecules has the chemical formula (C$_{l^2-1}$H$_{l+1}$)$_N$, where $l\ge 3$.
Such a $l$-kekulene nanospring has the shape of a spiral graphene nanoribbon, see Fig.~\ref{fig01} (a) and (c).
Unlike the graphene helicoid, the spiral nanoribbon has an inner channel, which makes it a softer structure.
A more detailed description of the structure of graphene nanospring is given in \cite{Savin2024}.

The deformation of nanosprings is modeled using the force field described in Ref.~\cite{Savin2024}. This force field accounts for the deformation of valence bonds and angles, as well as torsional and dihedral angles, and van der Waals interactions between atoms~\cite{Cornell1995,Savin2010,Lisovenko2016,Pavlov2024,Ilgamov2024}.
Note that the results obtained do not depend on the type of force field used.
Thus, the AIREBO force field \cite{Stuart2000} used in the works \cite{Li2023,Zhan2018,Sharifian2022} will lead to the same results.

The Hamiltonian of a finite-length nanospring is given by
\begin{eqnarray}
H=\sum_{n=1}^N \frac12({\bf M}\dot{\bf X}_n,\dot{\bf X}_n)+V({\bf X}_{1},...,{\bf X}_N)
\nonumber\\
+\sum_{n=1}^{N-3}\sum_{k=n+3}^N W({\bf X}_n,{\bf X}_{k}), \label{f2}
\end{eqnarray}
where ${\bf X}_n=\{ {\bf x}_{n,j}\}_{j=1}^{N_{\rm C}}$ is the $3N_{\rm C}$-dimensional vector with the coordinates of the atoms of the $n$-th structural unit and ${\bf M}$ is the diagonal matrix of the masses of the atoms.
The first, second, and third terms on the right-hand side of the Hamiltonian~(\ref{f2}) represent the kinetic energy, the valence interaction energy, and the van der Waals interaction energy, respectively.

The van der Waals interactions are described by the Lennard-Jones potentials
\begin{eqnarray}
W({\bf X}_n,{\bf X}_{k})
=\sum_{j=1}^{N_{\rm C}}\sum_{i=1}^{N_{\rm C}} U_{LJ}(r_{n,j;k,i}),\label{f3}
\end{eqnarray}
where the distance between the $i$-th atom of the $k$-th structural unit and the $j$-atom of the $n$-th structural unit is $r_{n,j;k,i}=|{\bf x}_{k,i}-{\bf x}_{n,j}|$.
Here the (6,12) Lennard-Jones potential has the form
\begin{equation}
U_{LJ}(r)=\epsilon_c\{[(r_c/r)^6-1]^2-1\}, \label{f4}
\end{equation}
and the parameters are $\epsilon_c=0.002757$~eV, $r_c=3.807$~\AA~ \cite{Setton1996}.

To find the ground state of the nanospring, the following potential energy minimization problem is numerically solved using the conjugate gradient method \cite{Fletcher1964,Shanno1976}:
\begin{eqnarray}
Q&=&V({\bf X}_{1},...,{\bf X}_N)\nonumber\\
&&+\sum_{n=1}^{N-3}\sum_{k=n+3}^N W({\bf X}_n,{\bf X}_{k})
\rightarrow\min : \{ {\bf X}_n \}_{n=1}^N. \label{f5}
\end{eqnarray}
The ground states of the nanosprings of $N=180$ structural units (about 30 coils) are shown in Fig.~\ref{fig01}.

The following system of Langevin equations is integrated numerically to model the thermal oscillations of the nanosprings:
\begin{equation}
{\bf M}\ddot{\bf X}_n=-\frac{\partial H}{\partial {\bf X}}_n-\gamma{\bf M}\dot{\bf X}_n-\Xi_n,~n=1,...,N,
\label{f6}
\end{equation}
with the initial conditions corresponding to the ground state
\begin{equation}
{\bf X}_n(0)={\bf X}_n^0,~\dot{\bf X}_n(0)={\bf 0},~n=1,...,N, \label{f7}
\end{equation}
where ${\bf X}_n^0=\{\{(x_{n,j}^0,y_{n,j}^0,z_{n,j}^0)\}_{j=1}^{N_C}\}_{n=1}^N$ is the solution to problem Eq.~(\ref{f5}).
Here $\gamma=1/t_r$ is the friction coefficient characterizing the intensity of interaction of the nanospring with the Langevin thermostat (the relaxation time of particle velocity is $t_r=10$~ps),
$\Xi_n=\{(\xi_{n,j,1},\xi_{n,j,2},\xi_{n,j,3})\}_{j=1}^{N_{\rm C}}$ is the $3N_{\rm C}$-dimensional vector of normally distributed random Langevin forces with correlation functions
$$
\langle\xi_{n,l,i}(t_1)\xi_{k,m,j}(t_2)\rangle
=2M_{l}k_BT\gamma\delta_{nk}\delta_{lm}\delta_{ij}\delta(t_1-t_2),
$$
with $k_B$ being the Boltzmann constant and $T$ the thermostat temperature.

The equations of motion Eq. (\ref{f6}) are solved numerically using the velocity Verlet method \cite{Verlet1967}.
A time step of 1 fs is used in the simulations, since further reduction of the time
step has no appreciable effect on the results.

Once equilibrium is reached between the molecular system and the thermostat, the mean energy $\bar{E}$ and spring length $\bar{L}$ are determined.
The energy is given by the Hamiltonian (\ref{f2}), and the length is found as the distance between the centers of gravity of the first and last structural units.
The temperature dependencies of the energy and length of the nanosprings are then obtained.
Then the value of the dimensionless heat capacity coefficient is found $c(T)=(3NN_Ck_B)^{-1}d\bar{E}/dT$
and the axial thermal expansion coefficient of the nanospring $\alpha(T)=\bar{L}^{-1}d\bar{L}/dT$ is determined.
\begin{figure}[tb]
\begin{center}
\includegraphics[angle=0, width=1.0\linewidth]{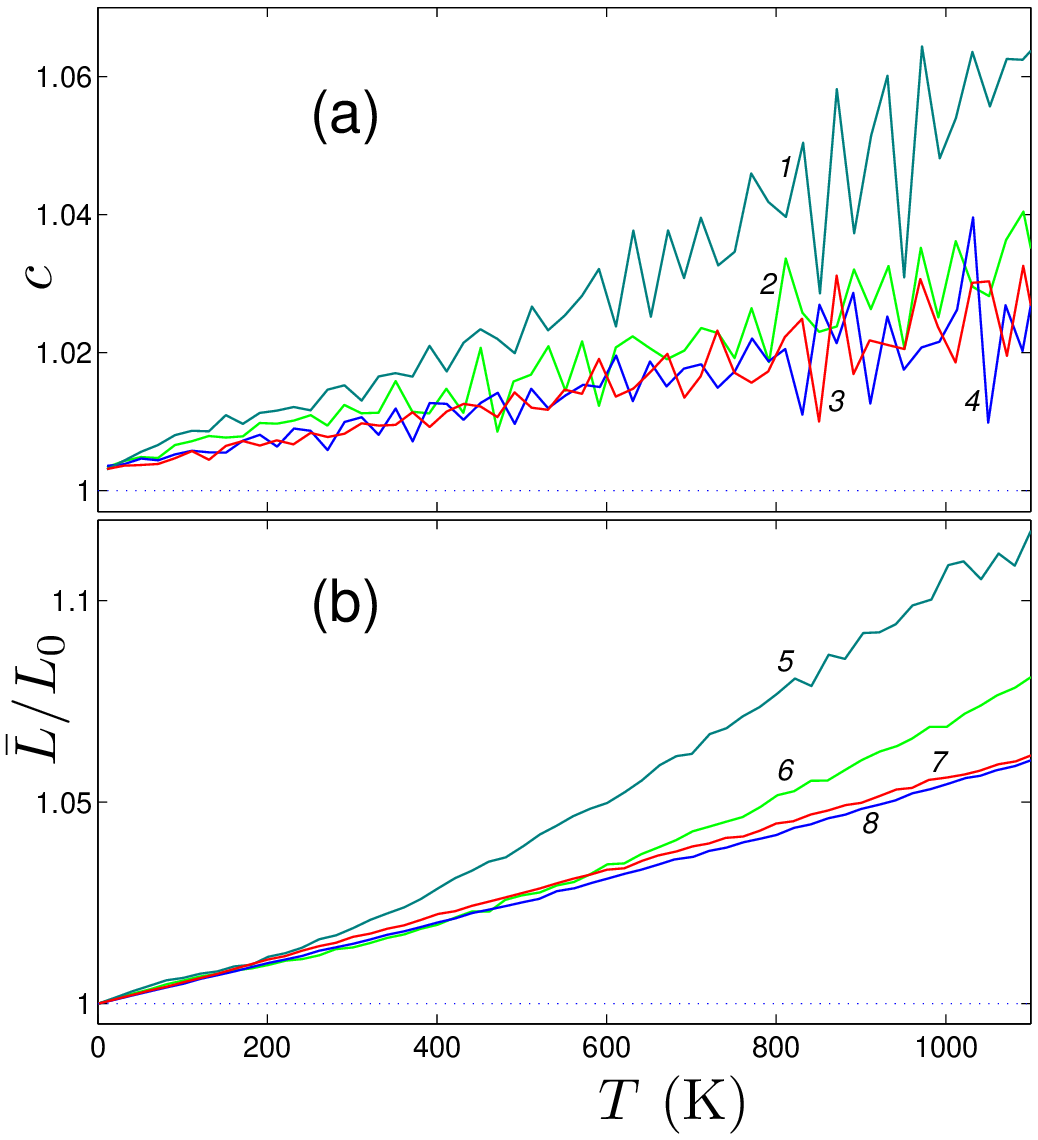}
\end{center}
\caption{\label{fig02}\protect
The temperature dependencies of (a) the dimensionless heat capacity $c$ and (b) the relative elongation $\bar{L}/L_0$ of the helical $l$-kekulene nanosprings (curves 1, 2 and 5, 6 are for $l=3$, 4, respectively) and the helical $l$-coronene nanosprings (curve 3, 4 and 7, 8 are for $l=3$, 4, respectively).
These nanosprings consist of $N=180$ structural units ($L_0$ is the length of the molecule in the ground state).
}
\end {figure}

Numerical modeling of nanosprings consisting of $N=180$ structural units (with a length of $L_0=10.4$~nm) has shown their stability to thermal fluctuations within a wide temperature range $0<T<1300$~K. 
An increase in temperature results in only a slight increase in the dimensionless heat capacity and an increase in helix length (see Fig.~\ref{fig02}). 
The coefficient of axial thermal expansion is $\alpha\approx 5\times 10^{-5}$~K$^{-1}$. 
The growth of the heat capacity and the increase in length are due to the soft anharmonicity of the van der Waals interactions between atoms (soft anharmonicity of the Lennard-Jones potential Eq.~(\ref{f4})).
Thermal fluctuations do not lead to the formation of stable defects in the helix. 
It will be shown that such defects can be created by deforming the nanosprings through longitudinal compression, bending, and twisting.

\section{ Axial compression of nanosprings \label{s3}}

The longitudinal compression of a nanosprings consisting of $N=180$ structural units is modeled. 
To do so, the ground state of the nanospring is used, and all the coordinates of the atoms in the first cell ($n=1$) and the $x$ and $y$ coordinates of the atoms in the last cell ($n=N$) are fixed. 
Conversely, the $z$ coordinates of the last cell decrease at a constant speed, bringing the ends of the helix closer together.
To achieve this, the system of equations of motion Eq.~(\ref{f6}) is numerically integrated with the boundary conditions
\begin{figure}[tb]
\begin{center}
\includegraphics[angle=0, width=1.0\linewidth]{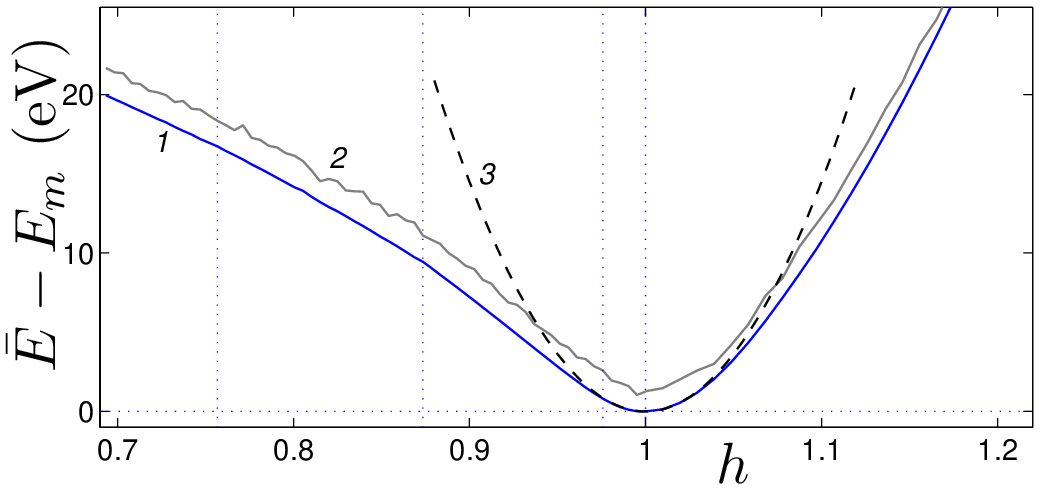}
\end{center}
\caption{\label{fig03}\protect
The dependence of the energy $\bar{E}-E_m$ of the 4-coronene nanospring (C$_{16}$H$_{4})_{180}$
on the relative longitudinal tension/compression $h$. 
Curve~1 shows the dependence at a temperature of 1~K, while curve~2 at a temperature of 300~K. 
Curve~3 corresponds to the harmonic nanospring with a stiffness coefficient of $K=4.4$~N/m. 
The vertical dotted lines show the characteristic values of relative compression: $h = 0.757$, 0.874, and 0.976.
}
\end{figure}
%
\begin{eqnarray}
{\bf X}_1\equiv {\bf X}_1^0,~x_{N,j}\equiv x_{N,j}^0,~y_{N,j}\equiv y_{N,j}^0,~\label{f8}\\
z_{N,j}(t)=z_{N,j}^0-vt,~j=1,...,N_{\rm C}, \nonumber  
\end{eqnarray}
and initial conditions Eq.~(\ref{f7}). 
The rate of compression is $v=0.05$~\AA/ps and the simulation temperature is $T=300$~K. 

After reaching the desired value of longitudinal dimensionless compression $h(t_0)=L(t_0)/L(0)$  at time $t = t_0$, the compression is stopped. 
Further modelling of the dynamics of the compressed nanospring with fixed edges is carried out by  numerically integrating the system of equations of motion Eq.~(\ref{f6}) with the boundary conditions
\begin{equation}
{\bf X}_1\equiv {\bf X}_1^0,~~{\bf X}_N\equiv {\bf X}_N(t_0), \label{f9}
\end{equation}
and the initial conditions 
\begin{equation}
{\bf X}_n(0)={\bf X}_n(t_0),~\dot{\bf X}_n(0)=\dot{\bf X}_n(t_0),~n=1,2,...,N.
\label{f10}
\end{equation}
After the system reaches the equilibrium with the thermostat, the mean value of its total energy $\bar{E}$ is found.
As the energy reference level it is convenient to take the energy of the ground state $E_m=E_0+3(N-2)N_{\rm C} Nk_BT$.
\begin{figure}[tb]
\begin{center}
\includegraphics[angle=0, width=1.0\linewidth]{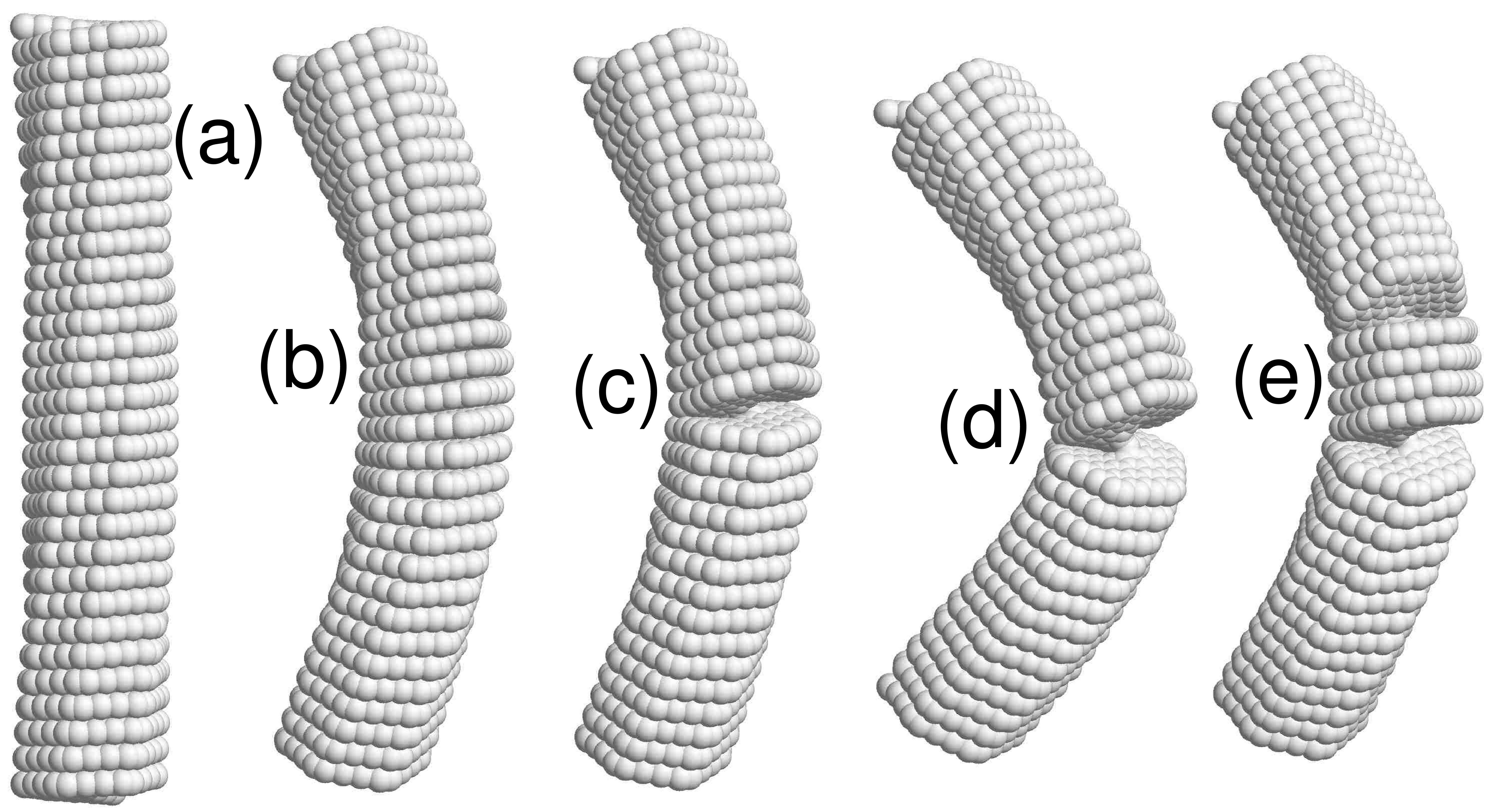}
\end{center}
\caption{\label{fig04}\protect
The $4$-coronene nanospring (C$_{16}$H$_{4})_{180}$ under axial compression: (a) $h=0.981$, (b) 0.873, (c) 0.869, (d) 0.757, and (e) 0.752.
}
\end{figure}
\begin{figure}[tb]
\begin{center}
\includegraphics[angle=0, width=1.0\linewidth]{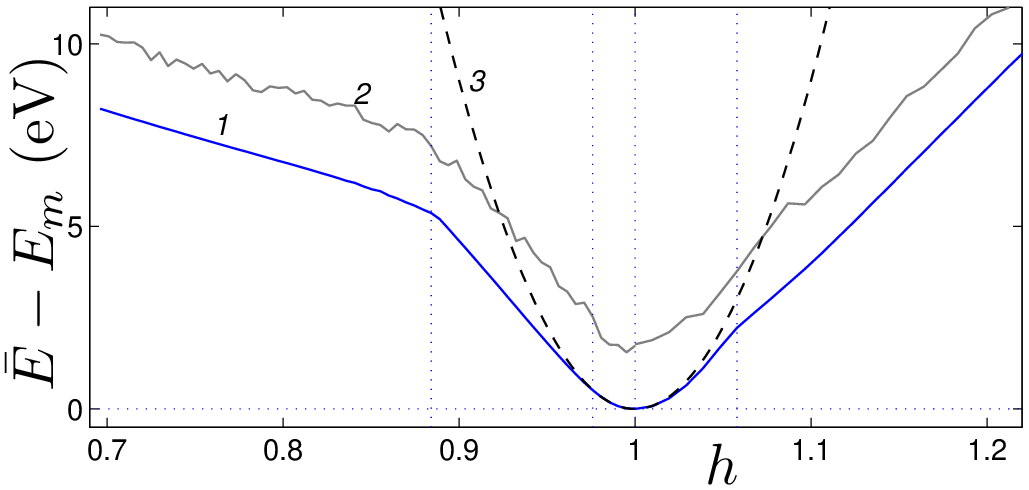}
\end{center}
\caption{\label{fig05}\protect
The dependence of the energy $\bar{E}-E_m$ of the 4-kekulene nanospring (C$_{15}$H$_{5})_{180}$ on the relative longitudinal tension/compression $h$. 
Curve~1 shows the dependence at a temperature of 1~K, while curve~2 at a temperature of 300~K. 
Curve~3 corresponds to the harmonic nanospring with a stiffness coefficient of $K=2.7$~N/m. 
The vertical dotted lines show the characteristic values of relative compression: $h = 0.884$, 0.976, and 1.058.
}
\end{figure}

The energy $\bar{E}(h)-E_m$ of the 4-coronene nanospring as the function of relative tension/compression $h$ is shown in Fig.~\ref{fig03}. Temperature has no significant effect on the shape of this function.
As can be seen, changing the temperature from 1~K to 300~K only results in a slight upward shift of the curve. The change in shape of the nanospring under compression is shown in Fig.~\ref{fig04}.

At weak relative compression, $1>h\ge h_1=0.976$, the nanospring energy grows proportionally to the parabola $(h-1)^2$. 
Within this compression range, the nanospring axis remains straight, see Fig.~\ref{fig04}(a). 
At $h=h_1$, the straight shape becomes unstable (Euler instability), and transverse bending occurs as the nanospring takes the form of a half-wave sinusoid, see Fig.~\ref{fig04}(b). 
Further compression increases the deviation of the nanospring axis from a straight line, and the energy growth of the compressed molecule follows an almost linear law. 
Thus, under axial compression, the nanospring behaves like a hinged rod. 
The half-wave sinusoidal shape loses stability at a relative compression of $h = h_2 = 0.874$. 
At this point, the smooth bending of the nanospring ends, and a transverse crack appears in the middle, see Fig.~\ref{fig04}(c). Further compression increases the crack opening. At $h = h_3=0.757$, a second crack appears in the nanospring, see Fig.~\ref{fig04}(d,e).
\begin{figure}[tb]
\begin{center}
\includegraphics[angle=0, width=1.0\linewidth]{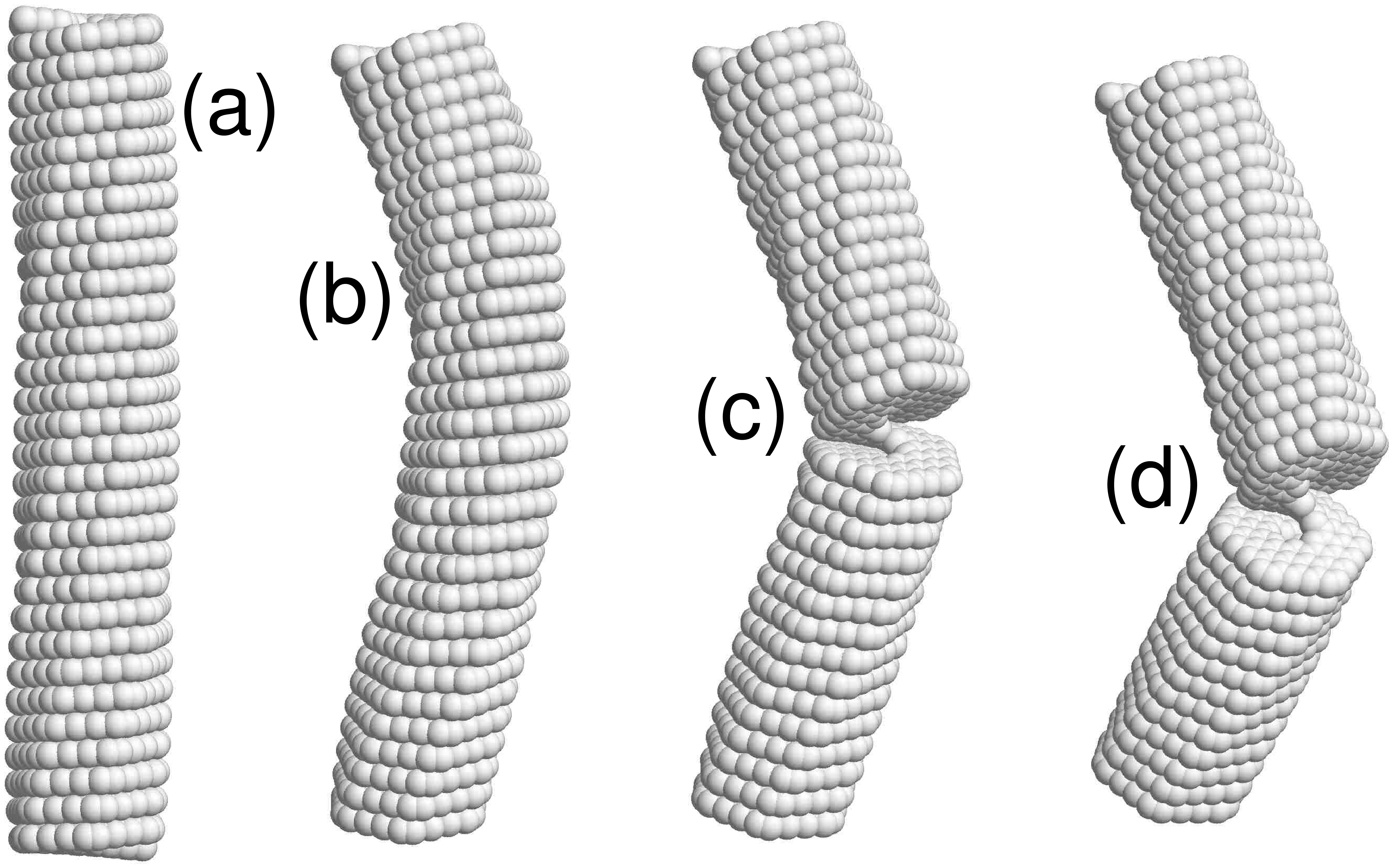}
\end{center}
\caption{\label{fig06}\protect
The $4$-kekulene nanospring (C$_{15}$H$_{5})_{180}$ under axial compression: (a) $h=0.981$, (b) 0.889, (c) 0.884, and (d) 0.797. 
}
\end{figure}

The energy $\bar{E}(h)-E_m$ of the $4$-kekulene nanospring (C$_{15}$H$_5$)$_{180}$ as the function of relative axial tension/compression $h$ is shown in Fig.~\ref{fig05}. 
The change in the shape of this nanospring under compression is shown in Fig.~\ref{fig06}. 
The nanospring behaves like a hinged Euler rod. 
At weak relative compression, $1>h\ge h_1=0.976$, the nanospring axis remains straight, and its energy grows quadratically, see Fig.~\ref{fig06}(a). 
At $h = h_1=0.976$, the straight configuration becomes unstable. 
The axis of the nanospring bends into the shape of a half-wave sinusoid, see Fig.~\ref{fig06}(b). 
Further compression increases bending and causes the energy of the compressed nanospring to increase almost linearly with $h$. 
Sinusoidal bending becomes unstable at $h=h_2=0.884$, at which point a transverse crack appears in the middle, see Fig.~\ref{fig06}(c). 
Further compression occurs as the crack opens wider, as shown in Fig.~\ref{fig06}(d). 

Note that, unlike the 4-coronene nanospring, compressing the 4-kekulene nanospring does not result in the formation of a second crack. Several cracks appear in the 4-coronene nanospring due to the presence of a rigid core that limits crack opening, making the formation of new cracks energetically preferable.
\begin{figure}[tb]
\begin{center}
\includegraphics[angle=0, width=1.0\linewidth]{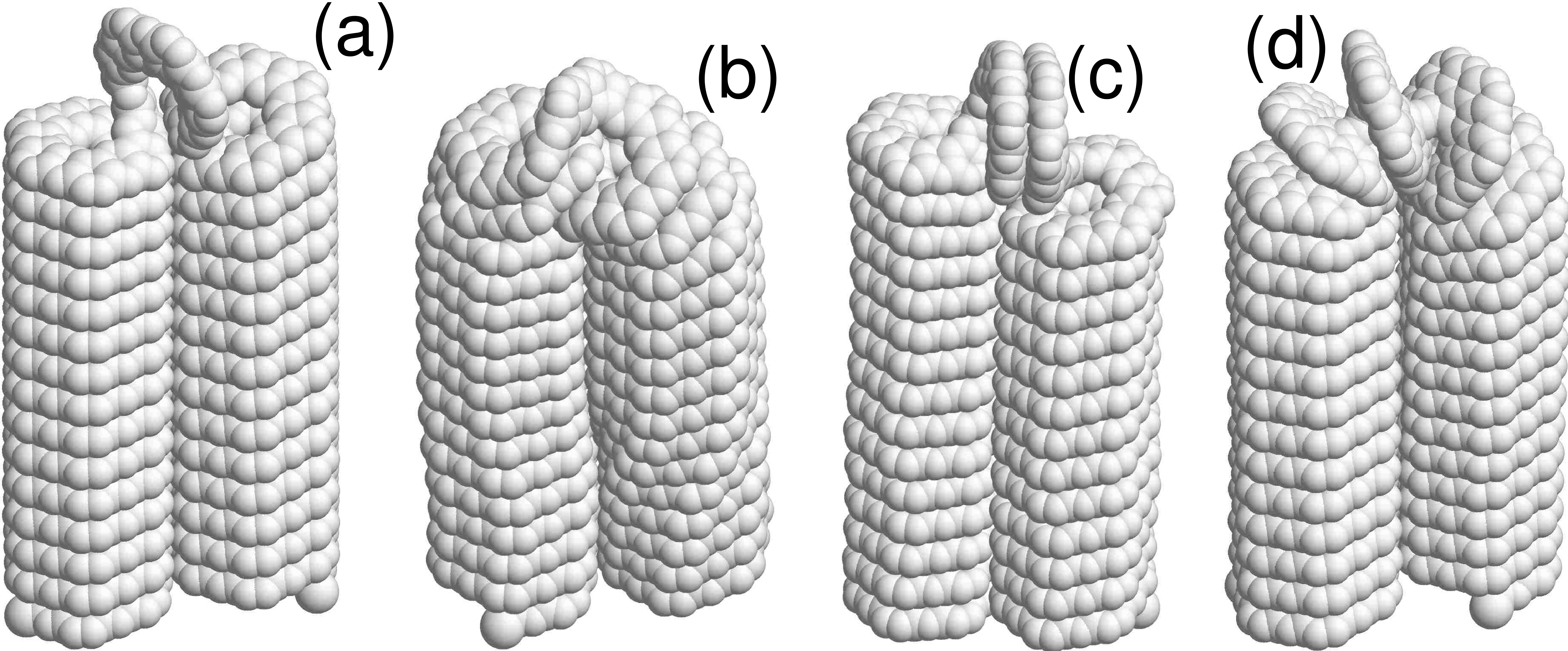}
\end{center}
\caption{\label{fig07}\protect
The equilibrium folded 3-kekulene nanosprings (C$_{8}$H$_{4})_{180}$ with energy (a) $E_d = 1.618$, (b) 1.629, and (c) 2.202~eV, and (d) the equilibrium folded $3$-coronene nanospring (C$_{9}$H$_{3})_{180}$
with energy $E_d=17.433$~eV. Energy is calculated relative to the ground state level, $E_d=E-E_0$.
}
\end{figure}

\section{Bending and fracture of nanosprings \label{s4}}

In order to bend the nanosprings, lateral forces must be applied in opposite directions to their ends and middle. 
This can be achieved by numerically integrating the following system of equations of motion
\begin{eqnarray}
{\bf M}\ddot{\bf X}_n&=&-\frac{\partial H}{\partial {\bf X}}_n-\gamma{\bf M}\dot{\bf X}_n-\Xi_n+Ff_n{\bf
e}_x, \label{f11} \\
&& n=1,2,...,N, \nonumber
\end{eqnarray}
where the index $n$ numbers the structural units; $F$ specifies the magnitude of the applied lateral force; the vector ${\bf e}_x=(1,0,0)$ specifies the direction along the $x$-axis; and the coefficients $f_1=f_N=1$, $f_{N/2}=f_{N/2+1}=-1$, and $f_n=0$ for the remaining values of $n$.
\begin{figure}[tb]
\begin{center}
\includegraphics[angle=0, width=1.0\linewidth]{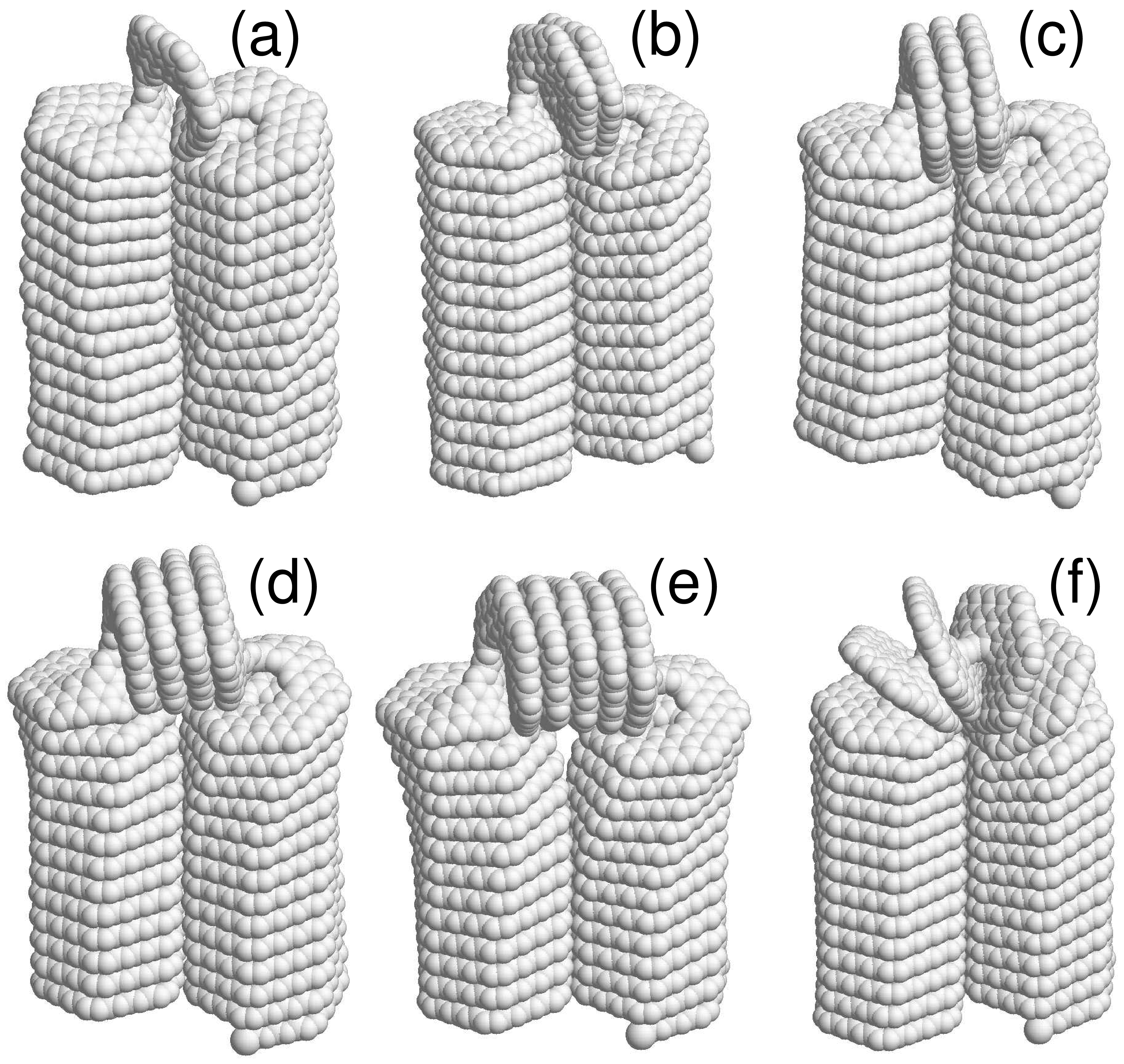}
\end{center}
\caption{\label{fig08}\protect
The equilibrium folded $4$-kekulene nanosprings (C$_{15}$H$_{5})_{180}$ with energy (a) $E_d=7.080$, (b) 8.072, (c) 8.628, (d) 9.428, (e) 10.247~eV, and (f) equilibrium folded 4-coronene nanospring 
(C$_{16}$H$_{4})_{180}$ with energy $E_d=34.337$~eV. The energy is counted from the ground state level, $E_d = E - E_0$.
}
\end{figure}

\begin{figure}[tb]
\begin{center}
\includegraphics[angle=0, width=0.78\linewidth]{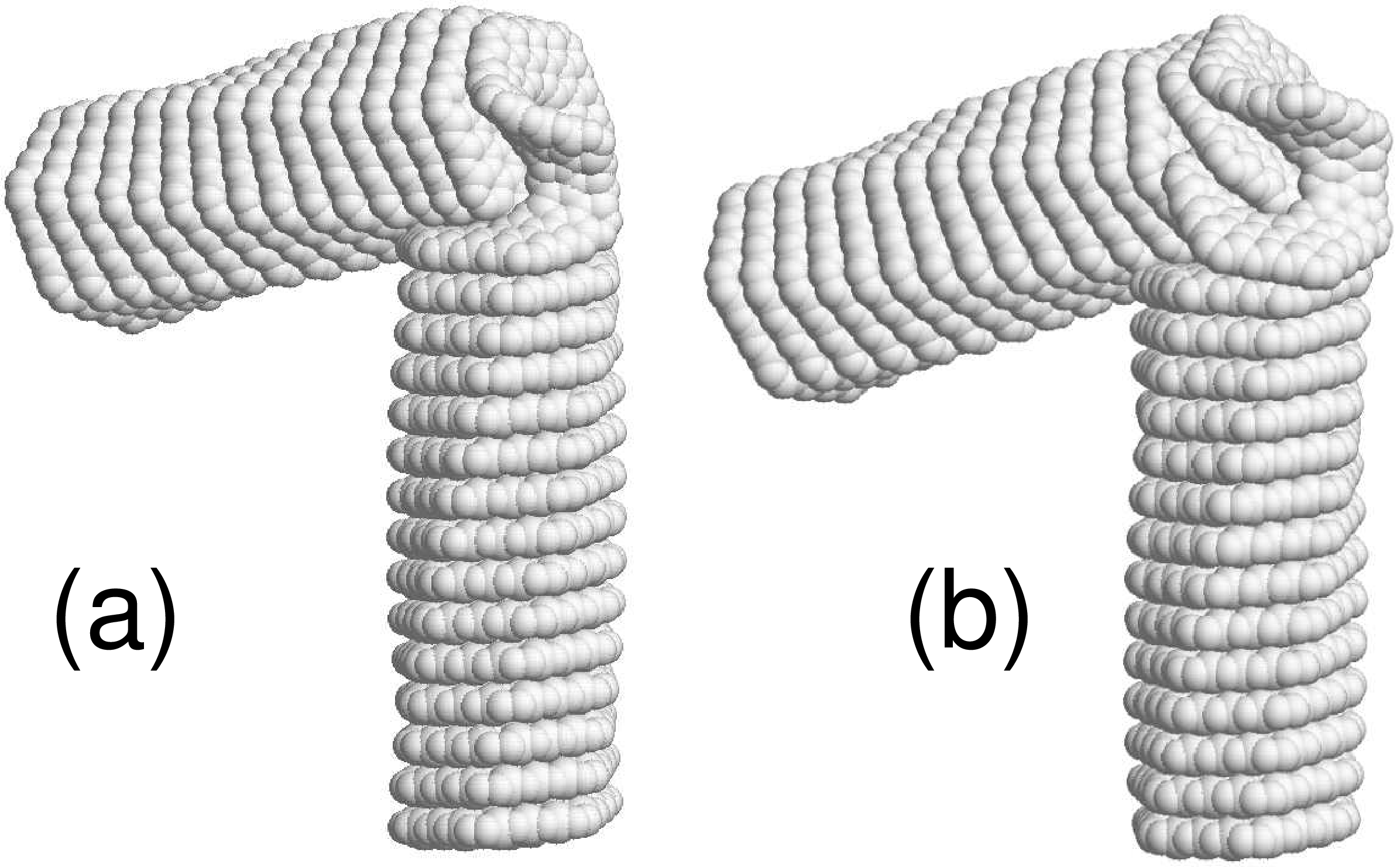}
\end{center}
\caption{\label{fig09}\protect
Stationary states of $4$-kekulene nanospring (C$_{15}$H$_{5})_{180}$ with fracture with energy (a) $E_d = 6.974$ and (b) 8.650~eV. The break angle is (a) $\phi_d=74^\circ$ and (b) $70^\circ$.
}
\end{figure}

Integrating the system of equations of motion Eq.~(\ref{f11}) with the initial conditions Eq.~(\ref{f7}) shows that there is a critical force, $F=F_0$, at which the nanospring experiences irrecoverable fracture.
When $F < F_0$, the nanospring takes the form of an arc under the action of lateral forces. 
When the load is removed by setting $F=0$, the nanospring returns to its original ground state. 
When the force is equal to the critical value $F = F_0$, the bending load leads to a fracture in the middle of the nanospring, and removing the load does not cause the nanospring to return to its initial ground state. After unloading, the nanospring can be in a folded steady state stabilized by van der Waals interactions between closely adjacent halves, see Figs.~\ref{fig07} and \ref{fig08}. 
Stable structures with a break angle of $\phi_d\approx 70^\circ$ are also possible, see Fig.~\ref{fig09}.
These structures are stabilized by topological defect formed at the corner.

For the 3-coronene nanospring (C$_9$H$_3)_{180}$, the critical force is $F_0=0.036$~eV/\AA, while for the 4-coronene nanospring (C$_{16}$H$_4)_{180}$ value $F_0=0.034$~eV/\AA. 
For $F>F_0$, helix bending can only lead to the formation of stable folded states, see Fig. \ref{fig07} (d) and \ref{fig08} (f). 
It can be seen that folding of the 3-coronene nanospring occurs by opening three coils, and folding of the 4-coronene nanospring occurs by opening four coils.

For the 3-kekulene nanospring (C$_8$H$_4)_{180}$, the critical force value is $F_0=0.006$~eV/\AA, while for the 4-kekulene nanospring (C$_{15}$H$_5)_{180}$ it is $F_0=0.012$~eV/\AA. 
Here one can talk about two closely located 90 degree cracks separated by one or a few coils, see Fig.~\ref{fig07}(a-c) and Fig.~\ref{fig08}(a-e). 
All folded structures are stable, but the most energetically favorable is the folding of the helix with one coil separating the two cracks, see Fig.~\ref{fig07}(a) and Fig.~\ref{fig08}(a). 
The 4-kekulene nanospring can also form stable structures with a break of angle $\phi_d\approx 70^\circ$, see Fig.~\ref{fig09}. 

As mentioned above, the folded nanosprings shown in  Fig.~\ref{fig07} and Fig.~\ref{fig08} are stabilized by van der Waals interactions between the adjacent halves. The energy of these interactions increases proportionally to the length of the nanospring, $L$. Therefore, very short nanosprings cannot remain in the folded state after unloading because the van der Waals energy is less than the elastic energy of bending. Conversely, for sufficiently long nanosprings, the folded structure is more favorable energetically than the straight configuration.
Sufficiently long nanosprings will fold and form a bundle  of adjacent parallel fragments of the same length, as often happens with $\alpha$-helical regions of protein molecules~\cite{Schafmeister1997,Liu2006}.

\section{Helix reversal defects in nanosprings \label{s5}} 

\begin{figure}[tb]
\begin{center}
\includegraphics[angle=0, width=0.98\linewidth]{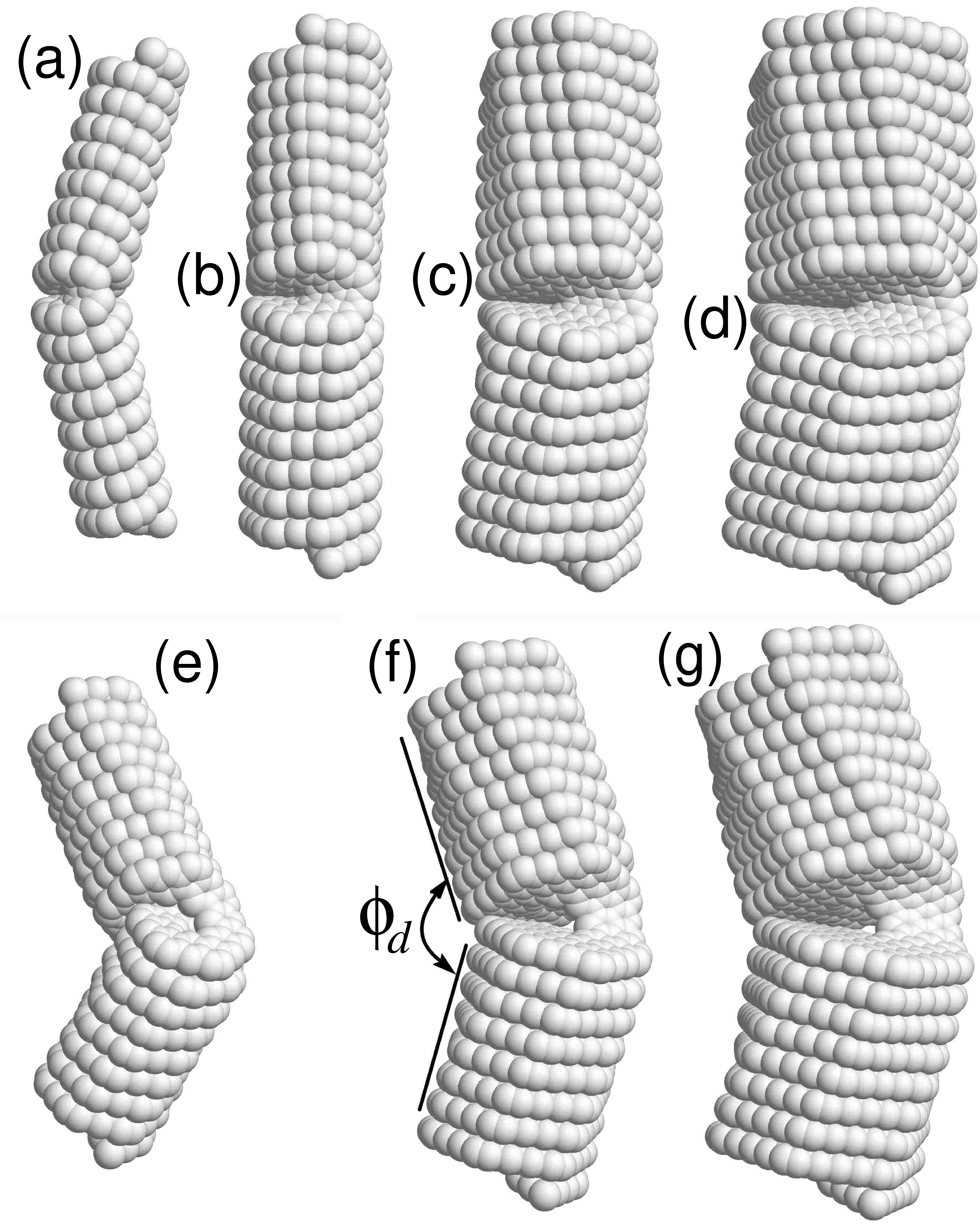}
\end{center}
\caption{\label{fig10}\protect
The helix reversal defects in $l$-coronene nanospring (in graphene helicoid) with (a) $l=2$, (b) $l=3$, (c) $l=4$, and (d) $l=5$, and in $l$-kekulene nanospring (in spiral graphene nanoribbon) with (e) $l=3$, (f) $l=4$, and (g) $l=5$. The angle between the axes of the two halves of the nanospring separated by the defect is denoted as $\phi_d$.
}
\end{figure}

Nanosprings in the form of helical macromolecules can exist in two equivalent ground states: a right-twisted helix or a left-twisted helix. An orientation defect occurs when one part of the macromolecule is a left-twisted helix and the other part is a right-twisted helix. 
This defect occurs at the boundary between these two regions, see Fig.~\ref{fig10}.
This structural defect describes a local change in the direction of rotation of the helix. 
Such defects are characteristic  of helical polymer molecules. 
Helix reversal defects are present in polytetrafluoroethylene (PTFE) crystals, where they cause helical inversion~\cite{Holt1999}, and in other helical polymers~\cite{Yashima2009,Rey-Tarrio2023,Jeon2025}.
\begin{table}[t]
\caption{The helix reversal defect energy $E_d$ and the angle $\phi_d$ between the axes of the two halves of the nanospring separated by the defect in the $l$-coronene nanospring with $l = 2, ..., 5$ and $l$-kekulene nanospring with $l = 3, 4,$ and 5.
\label{tab1}
}
\begin{center}
\begin{tabular}{ccccc|ccc}
\hline
 $l$  & 2 & 3  & 4 & 5 & 3 & 4 & 5\\
 \hline
$E_d$ (eV)         &  2.2  & 5.0  & 8.5   & 12.4  & 1.8  & 4.8   & 8.5 \\
$\phi_d$ ($^\circ$)& 136.7 & 146.3& 155.2 & 164.8 & 96.6 & 136.6 & 144.8\\
 \hline
\end{tabular}
\end{center}
\end{table}

Solving the minimum potential energy problem Eq.~(\ref{f5}) shows that the helix reversal defect is localized on two coils of the nanospring. 
The defect is characterized by the energy $E_d = E_1 - E_0$, where $E_1$ is the energy of the stationary state of the nanospring with the defect and $E_0$ is the energy of the nanospring without the defect.
The angle between the axes of the two halves of the nanospring separated by the defect is denoted as $\phi_d$, see Fig.~\ref{fig10}. 
The values of $E_d$ and $\phi_d$ for different nanosprings are presented in Table~\ref{tab1}.

\section{Relaxation of a highly stretched nanospring \label{s6}}

The structural and helix reversal defects discussed above can form when nanosprings are rapidly relaxed after being stretched. 
To demonstrate this, the dynamics of a 4-kekulene nanospring (C$_{15}$H$_{17}$)$_{400}$  is modeled.
At time $t=0$, the stationary state of the stretched nanospring with a relative elongation $h=L/L_0=5.5$ is considered. 
In this uniformly stretched state, the nanospring has length $L=127.4$~nm, and the axial and angular translational steps are $\Delta z=3.19$~{\AA} and $\Delta\phi=71^\circ$, respectively. 
The number of helix coils is $N_\phi=(N-1)\Delta\phi/2\pi=78.7$. 
In the ground state ($h=1$), the nanospring has axial and angular steps $\Delta z_0=0.58$~{\AA} and $\Delta\phi_0=61^\circ$, respectively, a length of $L_0=(N-1)\Delta z_0=23.2$~nm, and 67.6 coils.

To model the relaxation, the dynamics of a nanospring with free ends is considered. 
For this purpose the system of equations of motion (\ref{f6}) with the initial conditions 
\begin{equation}
{\bf X}_n(0)={\bf X}_n^0,~~\dot{\bf X}_n(0)={\bf 0},~~n=1,2,...,N,
\label{f12}
\end{equation}
is numerically integrated, where the vector $\{ {\bf X}_n^0\}_{n=1}^N$ defines the stationary state of the initially stretched nanospring.
\begin{figure}[tb]
\begin{center}
\includegraphics[angle=0, width=1.0\linewidth]{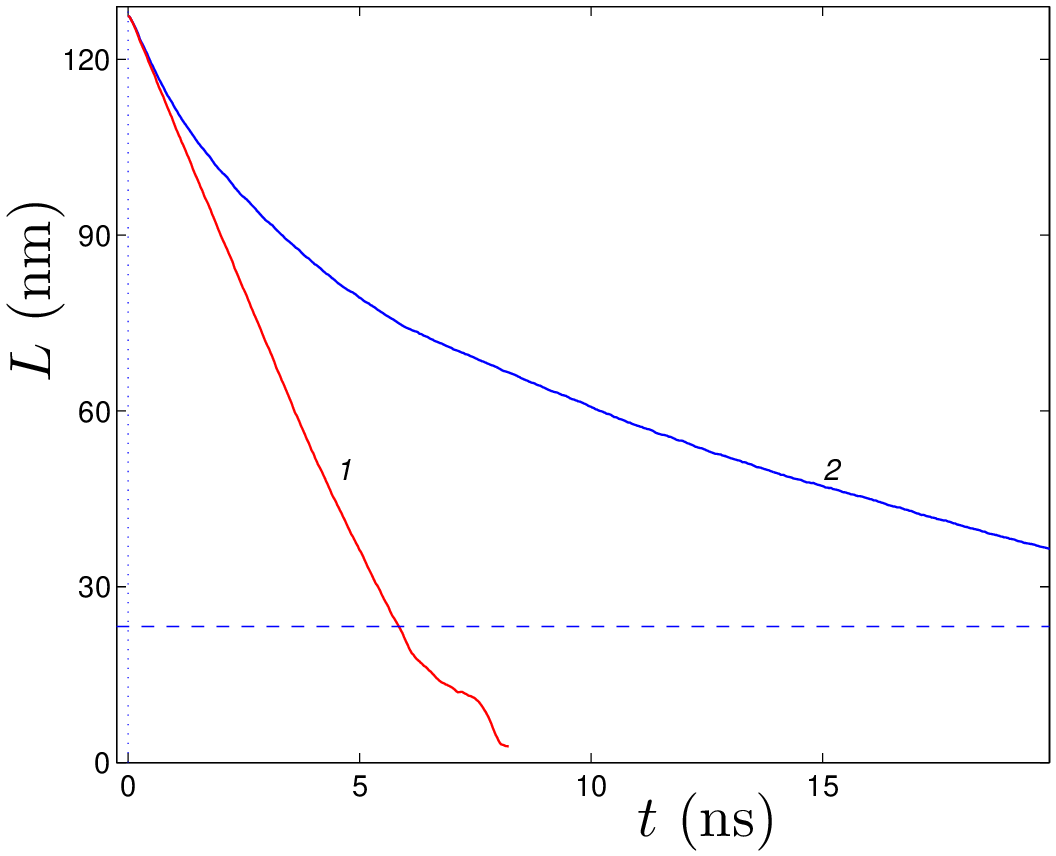}
\end{center}
\caption{\label{fig11}\protect
Relaxation of the 4-kekulene nanospring (C$_{15}$H$_{5}$)$_{400}$ initially stretched up to $h = 5.5$. The dependence of the nanospring length $L$ on time $t$ is shown. 
Curve~1 is obtained in the absence of interaction with the thermostat ($\gamma=0$, $T=0$), and curve~2 is obtained when the nanospring interacts with the thermostat ($\gamma=0.1$~ps$^{-1}$, $T=300$~K). 
The dotted line shows the value of the equilibrium nanospring length $L_0$.
}
\end{figure}

\begin{figure}[tb]
\begin{center}
\includegraphics[angle=0, width=0.68\linewidth]{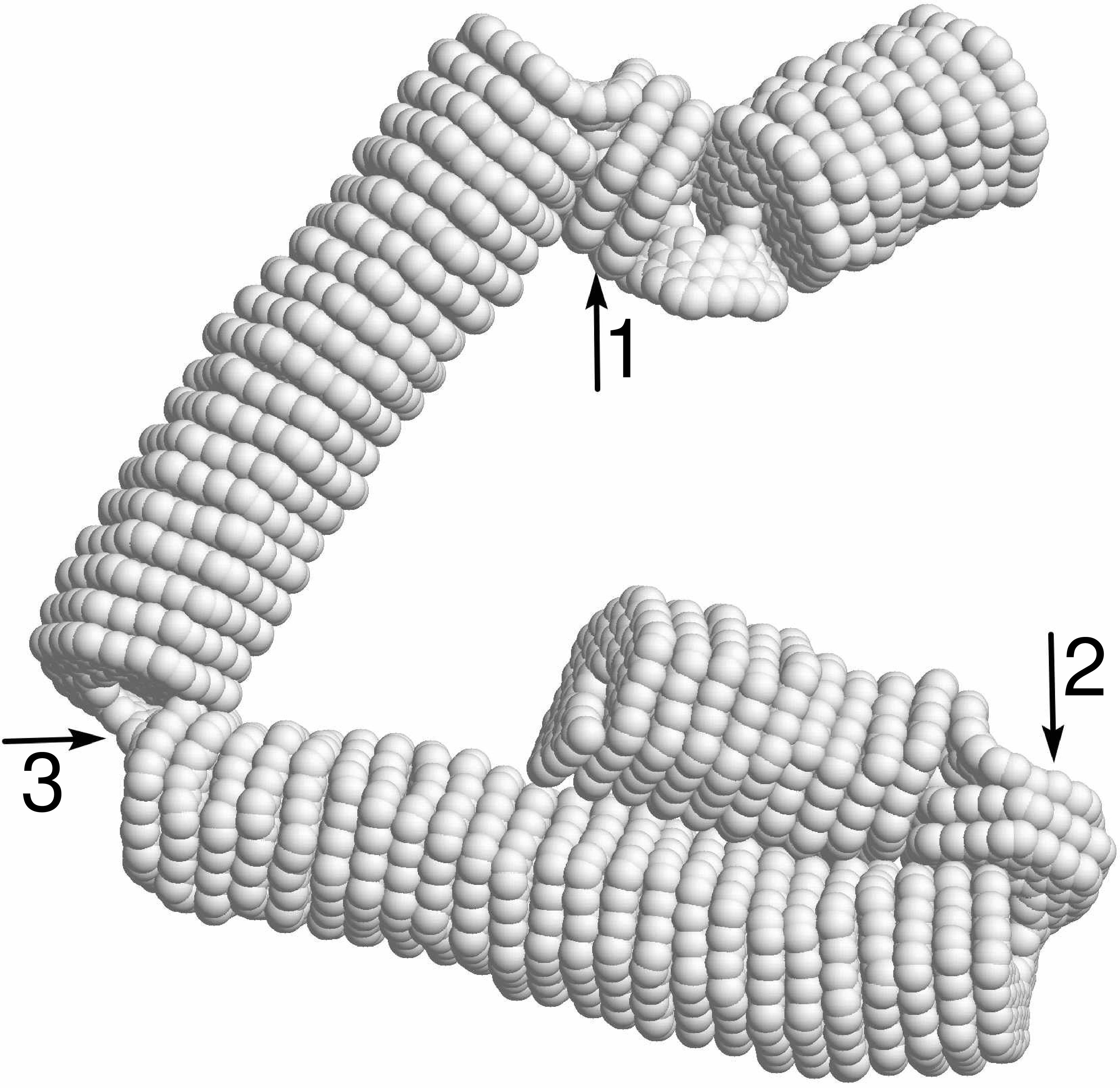}
\end{center}
\caption{\label{fig12}\protect
The structure of the initially stretched 4-kekulene nanospring (C$_{15}$H$_{5}$)$_{400}$ after relaxation.
The nanospring dynamics was simulated without considering the interaction with the thermostat. 
Arrows~1 and 2 show pairs of helix reversal defects, and arrow~3 shows the nanospring fracture.
}
\end{figure}

It is found that in the absence of interaction with the thermostat (at friction coefficient $\gamma=0$ and temperature $T=0$) the edges of the stretched nanospring converge with a constant velocity $v=1866$~m/s, see curve 1 in Fig.~\ref{fig11}. The convergence occurs due to the formation of growing non-stretched regions with longitudinal $\Delta z_0$ and angular pitch $\Delta\phi_0$ at the ends of the helix. Without rotation of these end sections, their convergence would lead to the formation of $N_{\phi}-N_{\phi,0}=11.6$ coils with opposite twist inside the nanospring leading to the formation of orientation defects. In the absence of interaction with the thermostat, a very rapid contraction of the stretched nanospring is accompanied by a relatively slow rotation of the ends, which prevents the negative twist in the center of the chain from being fully eliminated. Consequently, two sections with negative twist form in the nanospring, see Fig.~\ref{fig12}. 
Helix reversal defects form at the edges of these sections. In addition to these four defects, a structural defect (nanospring fracture) is formed.

If the relaxation of the stretched nanospring takes place in a viscous medium, i.e. taking into account its interaction with the thermostat, viscosity leads to slowing down of the convergence of the ends, see curve~2 in Fig.~\ref{fig11}. 
In this case, the convergence time is sufficient to remove the negative twist arising in the center of the nanospring due to the rotation of the ends. 
Therefore, the helix relaxes directly to its ground state and no defects are formed.

\section{Twisting of nanosprings \label{s7}}

The orientation defect in a nanospring can also be obtained by twisting, which will be simulated for a nanospring of $N = 200$ structural units starting with the ground state. The position of atoms of the first structural unit ($n = 1$) are fixed, and the last structural unit ($n=N$) is rotated around the nanospring axis with a constant angular velocity. The system of equations of motion (\ref{f6}) is numerically integrate with the following boundary and initial conditions
\begin{eqnarray}
&&{\bf X}_1(t)\equiv {\bf X}_1^0,~
\{{\bf X}_n(0)={\bf X}_n^0,\dot{\bf X}_n(0)={\bf 0}\}_{n=2}^{N-1}, \nonumber\\
&&x_{N,j}(t)=\cos(\omega t)x_{N,j}^0-\sin(\omega t)y_{N,j}^0,\label{f13}\\
&&y_{N,j}(t)=\sin(\omega t)x_{N,j}^0+\cos(\omega t)y_{N,j}^0,~j=1,...,N_{\rm C}\nonumber
\end{eqnarray}   
where $\omega$ defines the angular velocity of the last structural unit ($n=N$). 
The value $|\omega|=0.25$ ps$^{-1}$ is set.

The twisting of the nanospring starts at $t=0$ and ends at $t=t_0$, when the twist angle $\phi=\omega t_0$ is reached. 
Further modeling of the dynamics of the twisted nanospring with fixed values of $x$ and $y$ coordinates of the atoms of the last structural unit is carried out. 
After the system reached the equilibrium state, the average value of the total energy $\bar{E}$ is found, and then the twisting energy of the helix is found
$$
E_t=\bar{E}-E_0-(3N-5)N_{\rm C}k_BT,
$$
where $E_0$ is the ground state energy and $T$ is the thermostat temperature.
\begin{figure}[tb]
\begin{center}
\includegraphics[angle=0, width=1.0\linewidth]{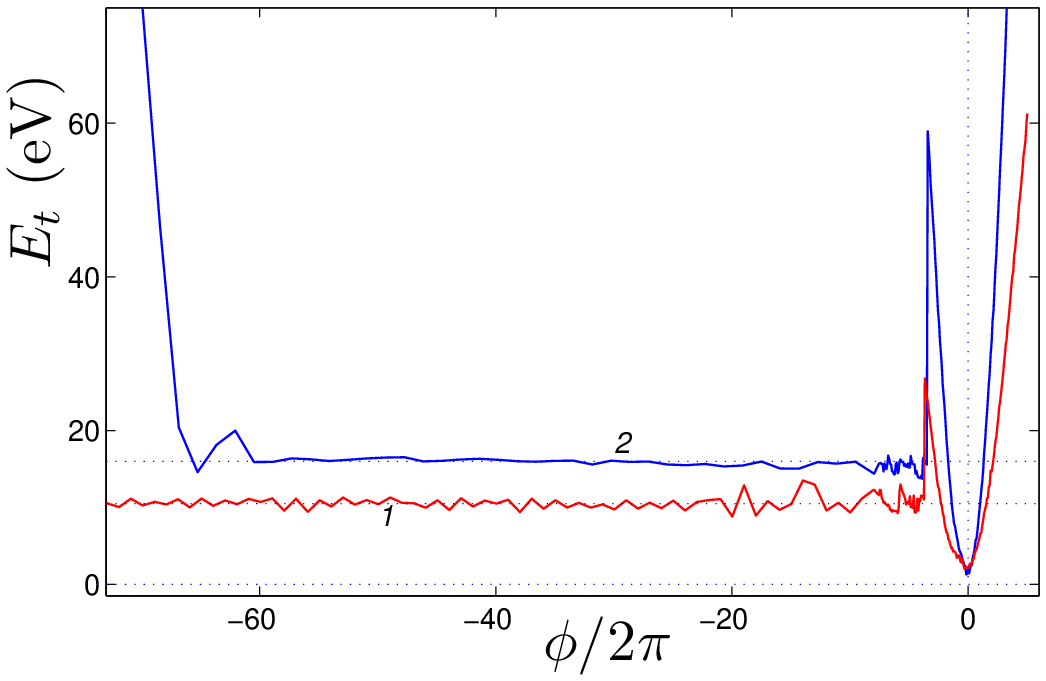}
\end{center}
\caption{\label{fig13}\protect
The energy $E_t$ of the twisted 4-kekulene (C$_{15}$H$_{5}$)$_{200}$ (curve~1) 
and 4-coronene (C$_{16}$H$_{4}$)$_{200}$ (curve~2) nanosprings  as the function of the twist angle around the spiral axis. 
The dotted horizontal lines correspond to values $E_t=10.5$ and 16~eV. 
The thermostat temperature is $T = 300$~K.
}
\end{figure}
\begin{figure*}[tb]
\begin{center}
\includegraphics[angle=0, width=0.9\linewidth]{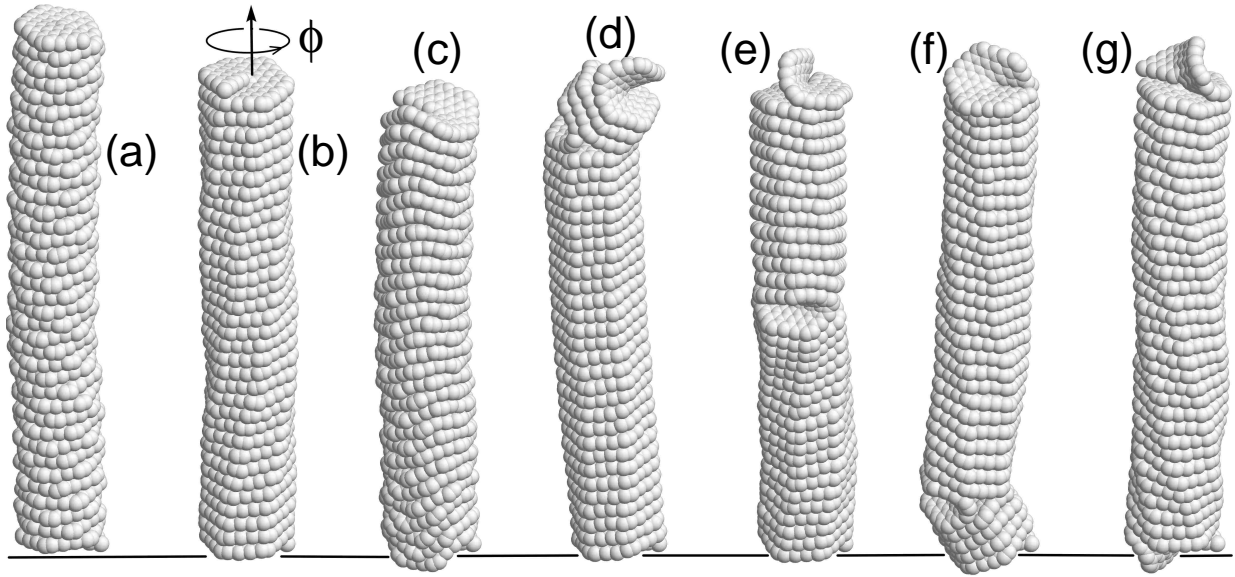}
\end{center}
\caption{\label{fig14}\protect
The structure of 4-coronene nanospring (C$_{16}$H$_{4}$)$_{200}$  under twist angle $\phi$: 
(a) $\phi=6.36\pi$, (b) 0, (c) -6.36$\pi$, (d) -12.74$\pi$, (e) -51.0$\pi$, (f) -111.4$\pi$, and (g) -130.4$\pi$. 
The horizontal line at the bottom shows the fixation of the atoms of the first structural unit.
}
\end{figure*}

The twist energy $E_t$ of the 4-kekulene and 4-coronene nanosprings as the function of the twist angle $\phi$ is shown in Fig.~\ref{fig13} by curves~1 and 2, respectively. 
For certainty, the nanosprings with a left-hand twist are taken; then for $\phi>0$ the twist increases, while for $\phi<0$ the nanospring is untwisted. 
When $\phi>0$, the additional twist occurs uniformly with gradual decrease in the angular pitch of the nanospring and an increase in the axial pitch, see Fig.~\ref{fig14}(a). 
In this case, the energy of the nanospring increases with twist angle proportionally to $\phi^2$.

When $\phi<0$, the nanospring untwisting is also uniform at the beginning. 
The angular pitch increases and the longitudinal pitch decreases, see Fig.~\ref{fig14}(c). 
The energy of the nanospring increases proportionally to $\phi^2$ until it reaches a maximum value at $\phi=\phi_0$ (for 4-coronene nanospring the critical value of the angle is $\phi_0\approx-7.2\pi$), after which it drops sharply. 
At this point an helix reversal defect forms in the nanospring. 
A part of the macromolecule near its upper end obtains the right-handed twist and the remaining part maintains the left-handed twist, see Fig.~\ref{fig14}(d). 
Further unwinding practically does not lead to a change in the nanospring energy so that the dependence $E_t(\phi)$ has a broad plateau, see Fig.~\ref{fig13}. 
Further rotation of the upper end of the nanospring leads to the movement of the helix reversal defect toward the lower end of the nanospring. 
After complete transition of the nanospring from left- to right-handed shape, see Fig.~\ref{fig14}(d-g), the energy $E_t$ starts to grow proportionally to $\phi^2$, see Fig.~\ref{fig13}.

\section{Conclusion}

A study was conducted to analyze the mechanical behavior of carbon nanosprings in the form of spiral macromolecules formed from $l$-kekulene and $l$-coronene molecules. 
This analysis was performed using molecular quasistatic (relaxational dynamics) and molecular dynamics simulations. 
The nanosprings were analyzed under axial compression, bending, and twisting. 
Earlier in the work~\cite{Savin2024}, the peculiarities of tensile deformation were investigated.

The primary distinction between $l$-kekulene and $l$-coronene nanosprings lies in the presence or absence of the inner channel. Specifically, $l$-kekulene features an inner channel, while $l$-coronene does not. This structural difference results in a distinct mechanical response to external forces.

The primary findings of the present study can be outlined as follows. 
\begin{enumerate}
\item 
The dimensionless heat capacity and the coefficient of axial thermal expansion were calculated in a wide range of temperatures, as shown in Fig.~\ref{fig02}. 
The heat capacity increases with temperature linearly due to the soft anharmonicity of the van der Waals interactions between coils of the nanosprings. 
The coefficient of axial thermal expansion is as large as $\alpha\approx 5\times 10^{-5}$~K$^{-1}$, which is significantly higher than that of many metals and alloys. 
This means that the use of carbon nanosprings in the production of temperature sensors is advantageous due to the thermal stability of $l$-kekulene and $l$-coronene molecules, which allows for sensors to function over a wide temperature range.
\item 
Nanosprings under axial compression have been shown to behave similarly to hinged elastic rods (see Fig.~\ref{fig04} for 4-coronene and Fig.~\ref{fig06} for 4-kekulene). 
They maintain a straight shape below the critical value of the compressive force, see panels (a), and demonstrate lateral buckling in the post-critical regime, see panels (b). 
It is evident from panels (c) that an even higher compressive force causes fracture of the nanosprings.
As illustrated in Fig.~\ref{fig04}(a), 4-coronene nanosprings lacking an inner channel may exhibit multiple cracks. 
In contrast, Fig.~\ref{fig06}(d) shows that for 4-kekulene, with an inner channel and consequently reduced bending stiffness, only a single crack is formed. 
\item 
The bending of nanosprings initiates with their arching, which is elastic deformation, and ceases once the bending forces are eliminated. 
At a certain level of bending force, nanosprings undergo irreversible changes in shape. In Fig.~\ref{fig07} and Fig.~\ref{fig08} the folded equilibrium structures of the 3-kekulene and 4-kekulene nanosprings are shown. 
These structures are stabilized by the van der Waals interactions between the halves of the folded nanosprings.
The folded structures exhibit even smaller potential energy than the straight nanosprings. 
In Fig.~\ref{fig09} another stable configuration of 4-kekulene nanospring is shown. 
This configuration is stabilized by the presence of a topological defect in the corner. 
This structure exhibits a higher potential energy than the straight nanospring.
\item 
Carbon nanosprings may exhibit helix reversal defects, separating the left-handed part from the right-handed part, as illustrated in Fig.~\ref{fig10}. 
The energies of the equilibrium helix reversal defects and the angle between the axis of the adjacent halves of the nanosprings are given for $l$-coronene and $l$-kekulene nanosprings in Tab.~\ref{tab1}. 
\item 
Twisting of nanosprings increasing its twist, leads to quadratic growth of potential energy with twist angle. 
Twisting in the opposite direction is more interesting. 
The potential energy increases quadratically at first, but after reaching a specific twist angle, the energy of the nanospring drops sharply. 
At this stage, an helix reversal defect is formed, and a part of the nanospring acquires the opposite chirality.
A subsequent twist leads to the movement of the helix reversal defect along the nanospring, ultimately resulting in a transformation of the entire structure to the opposite chirality. 
The structural transformation of the nanospring under twisting is illustrated in Fig.~\ref{fig14}.
\end{enumerate}

The results of this study show the unique behavior of carbon nanosprings when they are subjected to different types of deformation, such as compression, bending, and twisting. 
These deformation modes were not thoroughly explored in previous research. 
These results are particularly useful for the design of nanosensors that operate over a wide range of temperatures. 


\section*{Acknowledgements}

For A.V.S. the research work was funded by the Russian Science Foundation (RSF), grant No. 25-73-20038.

\section*{Data availability}
Data pertaining to this study are available upon request from the authors.

\bibliography{mybibfile}
\end{document}